# Energy dissipation and error probability in fault-tolerant binary switching

Mohammad Salehi Fashami[1], Jayasimha Atulasimha[1*] and Supriyo Bandyopadhyay[2]

[1]Department of Mechanical and Nuclear Engineering, [2]Department of Electrical and Computer Engineering, Virginia Commonwealth University, Richmond VA 23284, USA.

**Abstract**

The potential energy profile of a binary switch is a symmetric double well. Switching between the wells without energy dissipation requires time-modulating the potential barrier separating them and tilting the profile towards the desired well *at the precise juncture* when the barrier disappears. This demands perfect timing synchronization and is therefore fault-intolerant, even in the absence of noise. A fault-tolerant strategy that requires no time modulation of the barrier (and hence no timing synchronization) switches by tilting the profile by an amount at least equal to the barrier height and dissipates at least that amount of energy. Here, we present a third strategy that requires a time modulated barrier but *no timing synchronization*. It is therefore fault-tolerant in the absence of thermal noise and yet it dissipates arbitrarily small energy since an arbitrarily small tilt is required for slow and adiabatic switching. This case is exemplified with stress induced switching of a shape-anisotropic single-domain nanomagnet dipole coupled to a neighbor. We also show by examining various energy profiles and the corresponding probability distributions that when thermal noise is present, the minimum energy dissipated to switch in this scheme is $2kTln(1/p)$ [$p$ = switching error probability].

[*]*Corresponding author jatulasimha@vcu.edu*

The fundamental limits of energy dissipation in computing [1-7] are best understood by exploring the minimal energy dissipated to toggle a binary switch from one state to the other. The potential energy profile of such a switch is a *symmetric* double well as shown in the far left sketch of Fig. 1(a), with the two degenerate minima corresponding to the two stable states. One scheme for switching between the states [1, 2] entails modulating the potential barrier between the wells periodically in time. As the barrier is gradually eroded, the symmetric double well profile is first tilted towards the initial state to keep its potential energy constant. Just when the barrier is completely eroded and the well becomes monostable, it is translated horizontally in state space. The barrier then re-emerges on the opposite side of the well and therefore the system switches. This scheme results in vanishing dissipation because the system never acquires kinetic energy. However, it requires perfect timing synchronization between barrier modulation and the translation in order to switch accurately. That makes it fault-intolerant even in the absence of thermal noise.

Another scheme [3] that is dissipative but fault-tolerant is shown in Fig. 1(b). Here, the potential barrier is never modulated and hence no synchronization between two events is required. Whenever switching is desired, the potential profile is tilted towards the desired well in such a way that the tilt is at least equal to the maximum barrier height. This ensures that the system will definitely switch to the desired well in the absence of noise, but the system now gains kinetic energy equal to the amount of tilt, which is dissipated when the system relaxes to the new ground state (desired well). Clearly, error-resilience has been purchased with dissipation – a trade-off that is well-known in the context of the Fredkin billiard ball model of computation [7].

In this letter, we propose a new scheme that captures the best of both worlds. The barrier is modulated in time, but an arbitrarily small tilt towards the final state is maintained in the potential profile *all the time* as shown in Fig. 1(c). At the instant when the barrier disappears, the system automatically tends to switch to the final state, which is the minimum energy state, with some probability *p*. We will show that at a finite temperature, the energy $E_d$ dissipated in switching approaches $\sim 2kTln(1/p)$, based on equilibrium probability distribution prior to restoration of the barrier. The advantage of this (third) scheme over the first (dissipationless but error-prone) scheme is *error resilience without energy* *dissipation* at $T \to 0\,\text{K}$, and the advantage over the second scheme (dissipative but error-resilient) is the much lower energy-dissipation ($E_d \ll E_{barrier}$) without any additional error vulnerability at $T \to 0\,\text{K}$.

The third scheme that we propose here is also more practical to implement in nanomagnetic logic (NML) built with nanomagnetic binary switches [8] than the first. This is because NML chains consist of a linear array of nanomagnets where the first nanomagnet's state is propagated through all ensuing magnets by Bennett clocking [9]. Thus, the first magnet's state is set *before* the energy barrier separating the two stable states of the second magnet is modulated. Hence*, the "tilt" determined by the dipolar influence of the first magnet on the second is fixed and cannot be varied synchronously with the raising/lowering of the barrier in the second magnet* (see Fig. 2(b)).

We study this third scheme using a model nanomagnetic system both in the absence (T =0K) and presence (*T* = 4.2 K, 77K and 300 K) of thermal noise. The bistable switch chosen is a single-domain shape-anisotropic magnetostrictive nanomagnet shown in Fig 2 (a) which is elastically coupled to a piezoelectric layer of thickness 40 nm, forming a 2-phase multiferroic. The magnetostrictive nanomagnet is shaped like an elliptical cylinder of dimensions ~ 105 nm × 95 nm × 5.8 nm. The two stable states of the magnetization vector are along the major axis of the ellipse which is the easy axis of magnetization.

Application of a voltage across the piezoelectric layer generates uniaxial mechanical stress along the major axis via the $d_{31}$ coupling in the piezoelectric material if mechanical clamps are placed on the magnet's sides to prevent expansion/contraction along the minor axis. This uniaxial stress can rotate the magnetization vector provided the product of the magnet's magnetostrictive constant and stress (compressive stress is negative and tensile positive) is negative [10]. We choose this system simply because it is precisely illustrative of the switching methodology proposed here.

We will consider a 2-magnet system where the line joining their common center is aligned along the minor axis of the ellipse. This is shown in Fig. 2(b). As long as the separation between the magnets is not too small, the dipole coupling between the magnets will favor anti-ferromagnetic ordering where the magnetizations of the two magnets will be mutually anti-parallel. Assume now that the *left magnet is a hard magnet (larger shape anisotropy)* which is permanently magnetized in the down direction. Also assume that an external agent had switched the magnetization of the right (soft) magnet down, thus making the ordering temporarily ferromagnetic, which is an excited state. One would expect that once the external agent is removed, the right magnet will spontaneously flip up to allow the system to relax to the ground state, but this may not happen. This is because the magnetization of the right magnet has to overcome an energy barrier caused by its own anisotropic shape before it can flip up. An applied uniaxial stress along the easy axis can depress the shape anisotropy energy barrier and make the switching possible. *Therefore, the right magnet is a physical embodiment of the switching scheme shown in Fig. 1(c).* The energy barrier between its two stable states (magnetization orientations "up" and "down") is modulated by stress, and the "tilt" is due to the dipole coupling with the left magnet.

The magnetization dynamics of the second magnet under stress is studied in the manner of Ref 11, 12, 13 (based on solution of the Landau-Lifshitz-Gilbert equation) and is described in the supplementary material accompanying this Letter. When a critical stress of 3.145 MPa (induced by a voltage of only ~15 mV across the piezoelectric layer) is generated in the second magnet, the stress anisotropy energy equals the shape anisotropy energy barrier. This completely erodes the shape anisotropy energy barrier of ~ 0.75 eV separating the two stable magnetization states along the major axis of the ellipse. Thus, at 0 K (or in the absence of thermal fluctuation), an arbitrarily small dipole interaction energy due to the first magnet, that tilts the potential profile of the second magnet barely, is sufficient to switch the second magnet to the desired state, given sufficient time. This can be seen in Fig 2(c), which shows the energy dissipated to switch the second magnet *with 100% probability.* This energy dissipation decreases with increasing spacing between the magnets, or decreasing dipole coupling (tilt). Clearly, the switching is error-resilient since it occurs with 100% probability and yet the accompanying dissipation can be made arbitrarily small (say ~$3.2 \times 10^{-21}$ J/bit or ~ 0.02 eV) by increasing the inter-magnet separation and thereby decreasing the dipole coupling energy which is the tilt. A clearer picture emerges when we look at the nanomagnet's potential energy profiles as a function of its magnetization orientation for both strong (Fig 2 d) and weak (Fig 2 e) dipole coupling. Consider the case when the center-to-center separation between the magnets is R=150 nm as shown in Fig 2 (d). This separation causes a tilt of ~0.5 eV in the potential profile. When the barrier between the two states ~0.75 eV is removed by applying the critical stress, the magnetization rotates towards the correct ($\Phi$=90º) state favored by the dipole coupling or tilt. Thereafter, the barrier can be restored. No energy needs to be dissipated in modulating the barrier since the stress can be applied via a resonant resistance-inductance-capacitance circuit that is conservative [11]. However, energy is dissipated when the system switches and that energy is ~0.5 eV, which equals the tilt and has been verified by solving the LLG equations and estimating the energy dissipated through the Gilbert damping term (see supplement). If the dipole coupling is decreased by increasing the distance to R=400 nm, the tilt is reduced to a mere ~ 0.02 eV as shown in Fig 2 (e). Again when the ~0.75 eV barrier is removed with stress, the magnetization gradually moves closer towards the correct ($\Phi$=90º) state in ~100 ns (can be viewed as equivalent to a ball that would gradually move down a gentle slope). Upon completion of the switching, the barrier can be restored but the energy dissipated this time is a mere ~0.02 eV (equal to the tilt).

*Thus, in the absence of thermal fluctuations, the tilt and the resulting energy dissipation can be made vanishingly small, and yet switching always takes place without requiring any synchronization between the barrier modulation and the initiation of the tilt.* It is also critical to apply no more than the critical stress so that the barrier is just removed (see Fig 3 (a)). The reason for this is that the barrier needs to be "eroded", but not "inverted". If the barrier is inverted to create a monostable state, energy would be dissipated in the process of the magnetization reaching this state during lowering of the barrier, and this dissipation is unnecessary.

Next, we consider the trade-off between energy dissipated and dynamic switching error probability in the presence of thermal fluctuation at room temperature as depicted in Fig 3 (a) for the optimum case when $\sigma = \sigma_c$. We can derive an analytical relationship between $E_d$ and the error probability $p$ under the following assumptions:

(i) The E vs. $\Phi$ relationship is linear when critical stress is applied (this assumption is relaxed later to incorporate a sinusoidal profile as the dipole effect contains a $\sin(\Phi)$ term which does not yield a closed form solution for relations between $E_{dissipated}$ and $p_{error}$).

(ii) We apply the peak stress for long enough duration that it can be assumed that an equilibrium distribution (Boltzmann) is reached prior to withdrawal of stress. We further consider distribution with respect to $\Phi$ only and not $(\theta, \Phi)$ as we assume the magnetization vector at equilibrium is constrained to the plane of the magnet.

(iii) We also assume that when the stress is withdrawn suddenly (barrier restored), magnetizations that were in the $[-\pi/2, 0]$ half would settle to the "down" state while those in the $[0, \pi/2]$ half would settle to the "up" state. Extensive LLG analysis with thermal noise have shown that dynamic effects in restoring the barrier typically make $P_{error}$ larger than what would be estimated from this distribution. Nevertheless, this picture gives an estimate of the "minimum" energy that must be dissipated to limit the $p_{error}$ to a certain value. *Thus, the value of this analysis is to estimate a lower bound for energy dissipation which is certainly larger than the Landauer limit of $kTln(2)$ [1] which can be attained only with extremely complex time modulations of the barrier.*

**CASE I** Assuming that the energy of the down state ($\Phi = -90^0$) is $E_1$ and that of the up state ($\Phi = +90^0$) is $E_2$, ($E_1 > E_2$) and further assuming a linear relation between E and $\Phi$, one can write the probability distribution function in $\Phi$-space (assuming Boltzmann statistics) as:

$$\rho(\Phi) = Ae^{-\frac{(E_1-E_2)[\frac{\pi}{2}-\Phi]}{\pi kT}} \quad \text{where} \quad A = \frac{(E_1-E_2)}{kTL}(\frac{1}{e^{(E_1-E_2)/kT}-1}) \quad (1)$$

The probability that the magnetization is oriented between $\Phi = -90^0$ and $0^0$, just before the barrier is raised, is the error probability $p$ and can be found as:

$$p = \int_{-\pi/2}^{0} \rho(\Phi) d\Phi = \frac{e^{\frac{(E_{diss})}{2kT}} - 1}{e^{\frac{(E_{diss})}{kT}} - 1} \approx e^{\frac{-(E_{diss})}{2kT}} \quad \text{for } E_{diss} \geq 4kT \qquad (2)$$

where the energy dissipated is given by

$$E_{diss} = E_{tilt} = E_1 - E_2 \qquad (3)$$

Equation (2) can be recast as:

$$E_{diss} \approx 2kT \ln(1/p) \qquad (4)$$

Some of the trade-offs between dissipation and error probability were discussed in ref. [14, 15], but without deriving any analytical expression for the energy dissipated as a function of error probability by looking at magnetization distributions in $\phi$-space. We also differ from Ref 16 as we do not allow any energy recovery scheme. Note that Equation (4) is also counter-intuitive. Intuitively, one would expect that if Boltzmann statistics holds, then the relative probability of being in state $E_1$ (wrong state) with respect to $E_2$ (correct state) would be $e^{\frac{-(E_1-E_2)}{kT}}$ and hence the *static* error probability would be $e^{\frac{-(E_1-E_2)}{kT}}$ [6, 15] which would result in $E_{diss} = E_1 - E_2 = E_{tilt} = kT \ln(1/p_{static})$. Equating this with Equation (4), we get that $p = \sqrt{p_{static}}$ when we switch with critical stress. Thus, the dynamic error probability $p$ is greater than the static error probability $p_{static}$

**CASE II**

Here we incorporate a sinusoidal profile as the dipole effect contains a $\sin(\Phi)$ term. This represents a realistic energy profile for a multiferroic nanomagnet that is critically stressed so that the stress anisotropy exactly cancels the shape anisotropy. The probability distribution is now $\rho(\Phi) = Ae^{\frac{0.5(E_1-E_2)\sin(\phi)}{kT}}$ and has to be numerically integrated to find $A$ and $p_{error}$ unlike CASE I where an analytical result exists. When log(1/P$_{error}$) is plotted against E$_{dissipated}$=E$_{tilt}$ (See Fig 3 b) it can be approximated by an analytical result for E>0.1 eV (or 4 kT at room temperature)

$$\Delta E_{dissipated} / \Delta \log(1/p_{error}) = 1.9kT \qquad (5)$$

This estimate is slightly less conservative than (4) as the sinusoidal energy profile results in higher energy for the undesirable orientation (and less energy for the desired orientation) than a linear energy profile.

**CASE III**

Finally, we study the energy dissipation vs. dynamic error in a model nanomagnetic system shown in Fig 2 (b) by treating the second nanomagnet as a macro-spin [8]. We run extensive stochastic LLG analysis in the presence of thermal noise in the manner of Ref 17, by incorporating a random field due to thermal noise in the effective field term as described in the supplement. The results of this simulation are summarized in Fig 3 (b) where the $E_{dissipated}$ vs. $p_{error}$ is compared to the analytical estimate. The 3-D magnetization dynamics model typically shows higher $p_{error}$ for given $E_{dissipated}$ but the extent of deviation is more for higher $E_{dissipation}$ (or lower $p_{error}$). This is due to the out-of-plane distribution of magnetization as shown in Fig 3 (c) where being above the plane can help the switching while being below the plane can hurt the switching by producing a precessional torque that drives the magnetization to the wrong state. Thus, at a given temperature, increasing dipole coupling initially helps at low $E_{dissipated}$ (low tilt) but increasing the tilt does not decrease error as rapidly once the effect of this out-of-plane magnetization distribution dominates the switching error. Clearly, as the temperature increases, the out-of-plane distribution is more significant (large angles) and hence the switching error deviates from the analytical result (the decrease in $p_{error}$ with $E_{dissipated}$ saturates) at lower values of $E_{dissipated}$ or lower tilt (higher values of $p_{error}$). The in-plane and out-plane magnetization distributions just prior to restoring the barrier (withdrawal of stress) are discussed in detail in the supplement.

**IV. Conclusions**

In this Letter, we have shown with a concrete example that in the absence of thermal fluctuations, we can switch a binary switch with 100% probability and arbitrarily small energy dissipation while needing no timing synchronization in modulating the barrier. When thermal fluctuations are present, there is a trade-off between energy dissipated and the dynamic switching error probability. For a special case of switching a multiferroic magnet with critical stress, we have derived an analytical relationship between energy dissipated and the dynamic switching probability. We emphasize that for a practical logic switch *the minimum bound* for energy dissipated scales as *~2kTln(1/$p_{error}$)* and can therefore be larger than the Landauer limit of *kTln(2)* [1] which can be implemented only with extremely complex modulations of the barrier and precise timing sequences.

**Acknowledgement**

This work is supported by the US National Science Foundation under the SHF-Small grant CCF-1216614**,** NEB 2020 grant ECCS-1124714 and by the Semiconductor Research Corporation (SRC) under NRI Task 2203.001.

**Figure captions**

**Fig 1:** Switching stratergies involving:

(a) infinitely precise synchronization but near zero energy dissipation [2]

(b) no synchronization needed but energy dissipation greater than or equal to energy barrier [3]

(c) the new stratergy proposed: no synchronization and arbitrarily small energy dissipation at 0K .

**Fig 2:** (a) An elliptical multiferroic nanomagnet consisting of a piezoelectric layer in intimate contact with a magnetostrictive layer.

(b) A 2-magnet system comprising a hard magnet with large shape anisotropy and a soft multiferroic magnet with smaller shape anisotropy whose shape anisotropy energy barrier is modulated with stress.

(c) Energy dissipated in flipping (switching) the magnetization of the second nanomagnet as a function of the center-to-center separation between the two magnets (R).

(d) The energy profile of the multiforroic nanomagnet discussed in the text in the relaxed (unstressed) state and the critically stressed ($\sigma_{CR}$) state for large dipole coupling, R=150 nm

(e) The energy profile of the multiforroic nanomagnet discussed in the text in the relaxed (unstressed) state and the critically stressed state ($\sigma_{CR}$) for small dipole coupling, R=400 nm.

NOTE: $\sigma_{CR}$ ~ 3.1 MPa to exactly beat shape anisotropy but we need to chose a slightly higher $\sigma_{CR}$ ~ 3.15 to we have reasonable switching time to do our simulations.

**Fig 3:** Energy dissipated vs. minimum dynamic switching error

(a) Schematic that shows the effect of dipole coupling (tilt/asymmetry) and stress on the dynamic switching error. Energy profiles sketch at critical stress is shown. Blue thick line: energy profile, red dotted line: corresponding probability distribution function. Clearly, higher dipole coupling (tilt) will produce a more favorable probability distribution at the expense of dissipating more energy.

(b) Energy dissipated vs. dynamic switching error at various temperatures. The analytical bound is compared to estimates from 3-D simulation of magnetization dynamics.

(c) Schematic that shows the effect of spread in out-of-plane magnetization on the switching probability. In this particular case, out-of-plane orientation above the plane aids the switching whereas the out-of-plane orientation below the plane hurts the switching.

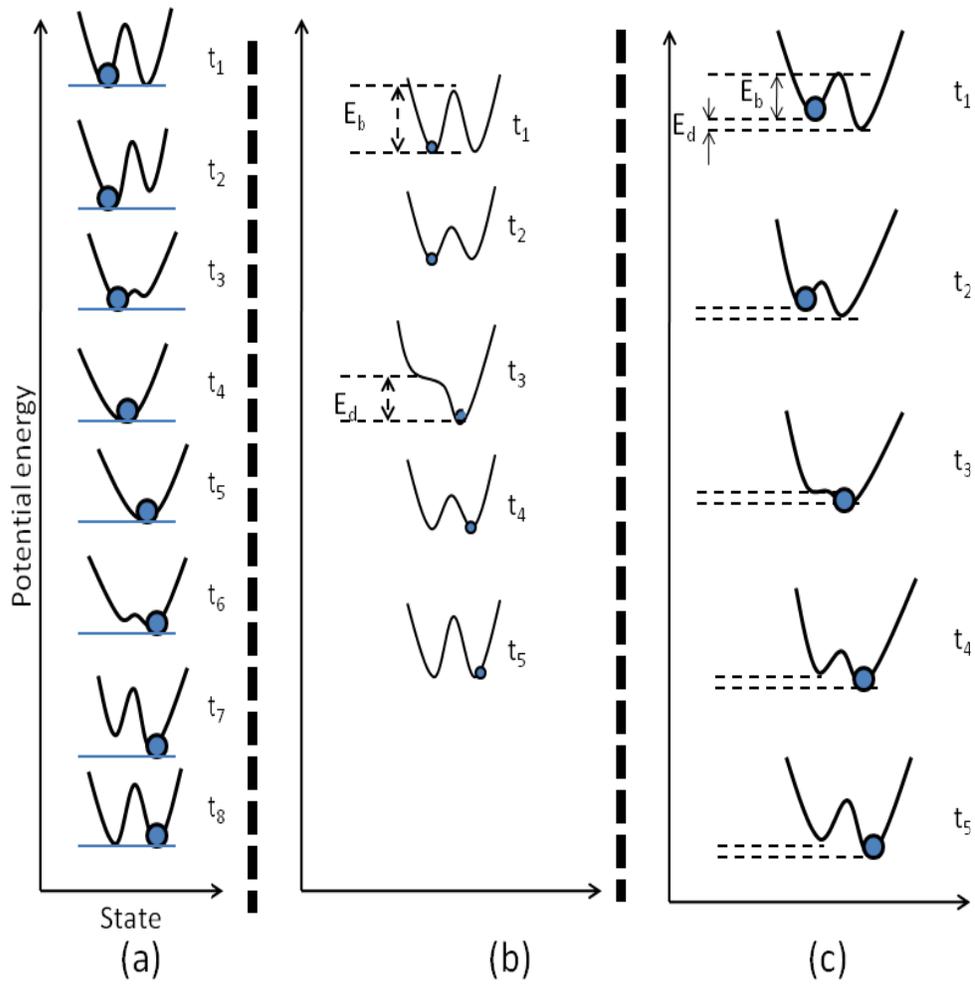

**Fig.1**

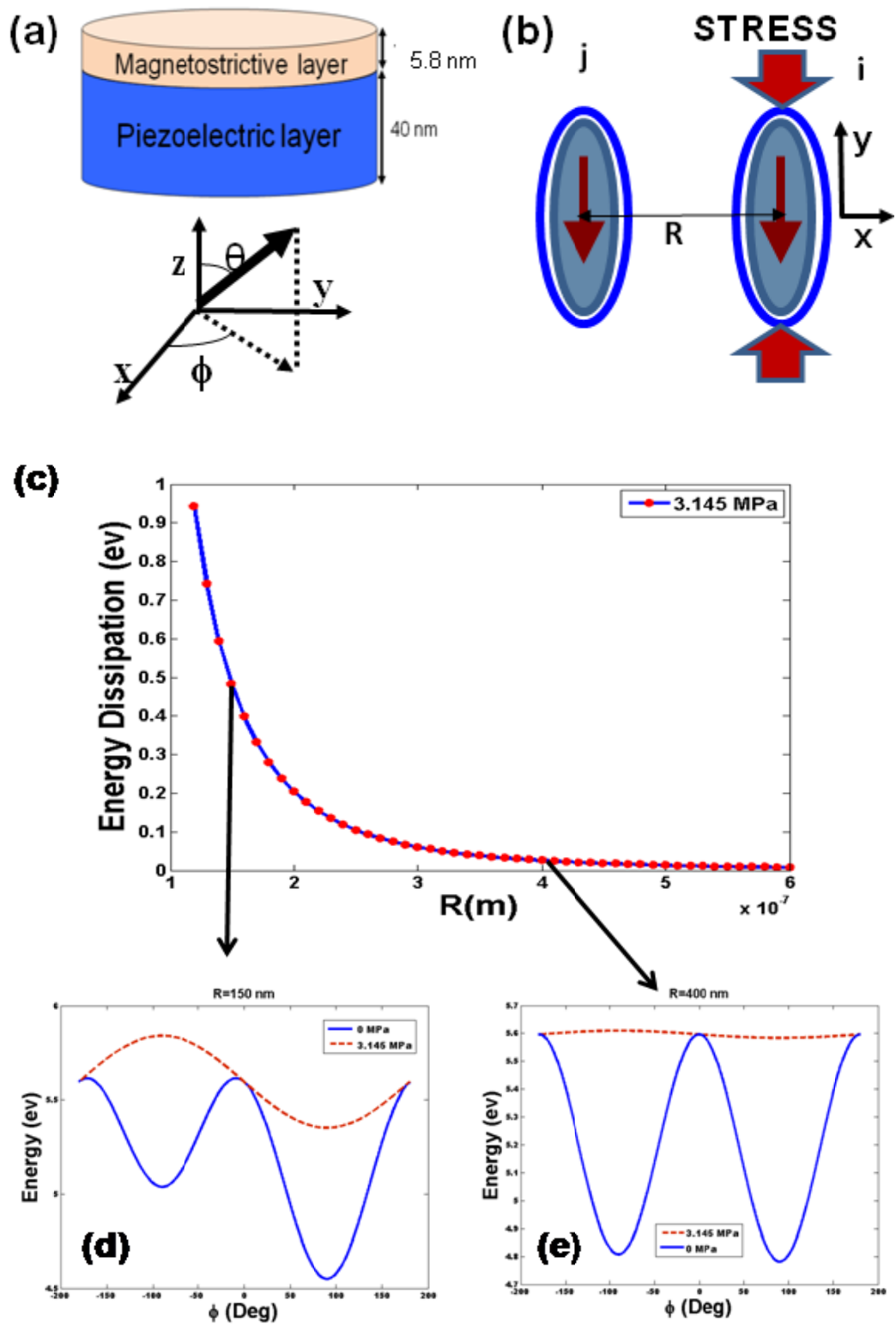

**Fig.2**

**(a)**

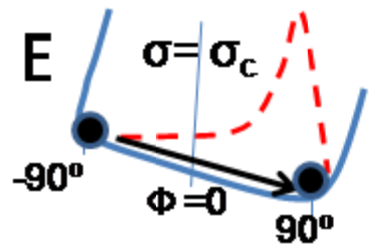

**(b)**

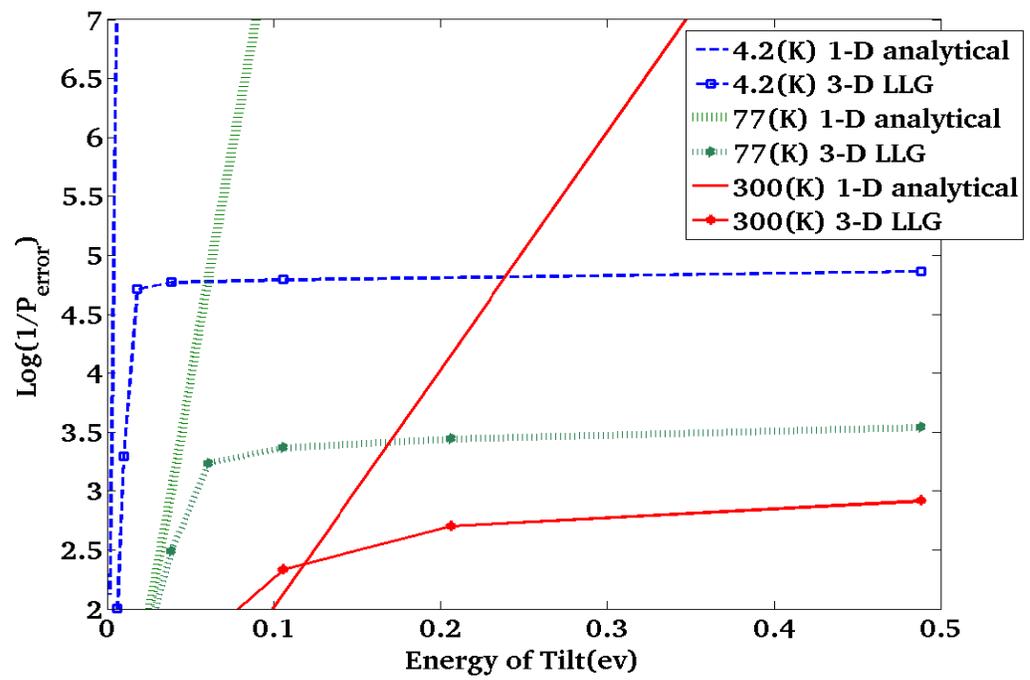

**(c)**

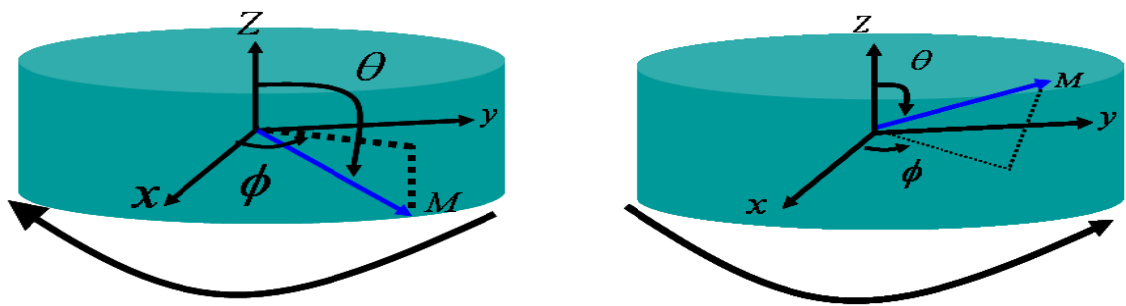

**Fig. 3**

**SUPPLEMENTARY MATERIAL FOR**

**Energy dissipation and error probability in fault-tolerant binary switching**

Mohammad Salehi Fashami[1], Jayasimha Atulasimha[1*] and Supriyo Bandyopadhyay[2]

[1]Department of Mechanical and Nuclear Engineering, [2]Department of Electrical and Computer Engineering, Virginia Commonwealth University, Richmond VA 23284, USA.

*jatulasimha@vcu.edu

In order to understand the magnetization dynamics described in the Letter, we solve the Landau-Lifshitz-Gilbert (LLG) equation adapted to model stress-induced magnetization dynamics in a magnetostrictive nanomagnet through the incorporation of stress anisotropy in the effective-field term. To define the "effective field", we first write the potential energy of the right magnet that is subjected to stress and interacts with its "stiff" left neighbor whose magnetization points down ($\Phi = -90^0$) via dipolar coupling as [R1, R7]:

$$U(t) = \underbrace{\left(\frac{\mu_0}{2}\right)\left[M_s^2\Omega\right]\left(N_{d\_xx}\left[\sin\theta(t)\cos\phi(t)\right]^2 + N_{d\_yy}\left[\sin\theta(t)\sin\phi(t)\right]^2 + N_{d\_zz}\left[\cos\theta(t)\right]^2\right)}_{E_{shape-anisotropy}}$$
$$\underbrace{-\left(\frac{3}{2}\lambda_s\sigma_i\Omega\right)\sin^2\theta(t)\sin^2\phi(t)}_{E_{stress-anisotropy}} - \underbrace{\frac{\mu_0}{4\pi R^3}M_s^2\Omega^2\left[\sin\theta_i\sin\phi_i\right]}_{E_{dipole}} \quad (S1)$$

where $\mu_0$ is the permittivity of free-space, $M_s$ is the saturation magnetization of the magnet, $\Omega$ is the magnet's volume, $N_{d\text{-}mm}$ is the demagnetization factor along the $m$-axis, $\lambda_s$ is the magnetostrictive coefficient of the magnet, $\sigma$ is the stress in the magnet, and $R$ is the center-to-center separation between the left and right magnets. The demagnetization factors are determined by the magnet's major and minor axes dimensions, as well as the thickness. The angles $\theta$ and $\phi$ are the polar and azimuthal angles of the magnetization vector in spherical coordinates (see Fig 2 a, b) of the Letter.

We assume that the magnetostrictive material is Terfenol-D ($M_s = 0.8 \times 10^6$ A/m, magnetostriction $\lambda_s = +900 \times 10^{-6}$ and Young's modulus (Y) $\sim 80 \times 10^9$ Pa [R2-R5]) and plot the potential energy $U$ versus $\phi$ in the plane $\theta = \pm 90^0$ (plane of the magnet) for zero stress and critical value of the stress ($\sigma_C$) in Fig. 2 (d) and Fig 2 (e) of the Letter. In the absence of dipole interaction, the potential profile would have been symmetric about $\phi = 0^0$ axis, with two degenerate minima occurring at $\phi = -90^0, \phi = +90^0$ in the absence of stress. The energy barrier separating these two minima would have been the shape-anisotropy energy barrier, with a magnitude of ~32 kT (at room temperature) for the choice of the material parameters and geometry made here. Note that compressive stress lowers the potential barrier while tensile stress raises it, because the product $\lambda_s \sigma$ is negative in the former case and positive in the latter. The potential profile in Fig. 2 (d) and Fig 2 (e) of the Letter replicate the profile for the switching scheme that is shown in Fig. 1(c) of the Letter. The "tilt" is due to dipole interaction, which can be varied by varying the spacing $R$ between the magnets.

If the system is initially in the left potential minimum (in Fig. 2 (d) and (e) of the Letter) i.e. magnetization pointing "down" along the major axis in Fig. 2 (b) of the Letter, then when we apply stress, we will depress the potential barrier, which will ultimately allow the system to switch from the left minimum at $\phi = -90^0$ (magnetization pointing "down") to the right minimum at $\phi = 90^0$ (magnetization "up") since the latter is slightly lower in energy because of dipole interaction. The energy difference between the left and right minima is the *minimum* internal energy dissipated in the magnet to switch in the presence of dipole interaction. In order to be minimally dissipative, switching must however be carried out adiabatically (very slowly). We exclude any dissipation in the external switching circuit, which can be made vanishingly small by switching adiabatically and recovering most of the energy expended in each clocking cycle using an LRC circuit [Main paper, reference 11].

The magnetization dynamics of a single domain nanomagnet under the influence of an effective field $\vec{H}_{eff}$ can be described by the Landau-Lifshitz -Gilbert (LLG) equation [R1, R6]:

$$\frac{d\vec{M}(t)}{dt} = -\frac{\nu}{1+\alpha^2}\vec{M}(t)\times\vec{H}_{eff}(t) - \frac{\alpha\nu}{M_s(1+\alpha^2)}\left[\vec{M}(t)\times\left(\vec{M}(t)\times\vec{H}_{eff}(t)\right)\right] \quad (S2).$$

Here $\vec{H}_{eff}^i$ is the effective magnetic field on the *i*-th nanomagnet in a linear chain of magnets, which is the partial derivative of its total potential energy ($U_i$) with respect to its magnetization ($\vec{M}_i$), $\nu$ is the gyromagnetic ratio, $M_s$ is the saturation magnetization of the magnetostrictive layer and $\alpha$ is the Gilbert damping factor [R6] associated with internal dissipation in the magnet when its magnetization rotates. Accordingly,

$$\vec{H}_{eff}^i(t) = -\frac{1}{\mu_0 \Omega}\frac{\partial U_i(t)}{\partial \vec{M}_i(t)} = -\frac{1}{\mu_0 M_s \Omega}\nabla_{\vec{m}} U_i(t) \quad (S3).$$

where $\Omega$ is the volume of any nanomagnet. The total potential energy the *i*-th element in the chain is given by equation (S1) and can be succinctly written as.

$$U_i(t) = E^i_{shape-anisotropy} + E^i_{stress-anisotropy} + E^{i-j}_{dipole} \quad (S4).$$

Here $E^{i-j}_{dipole-dipole}$ is the dipole-dipole interaction energy due to interaction between the stressed nanomagnet (i-th magnet) that acts as the bistable switch and its left neighbor (j-th magnet) that is assumed to be stiff and provides the input or tilt/asymmetry to switch the j-th magnet. This term is evaluated via the standard expression [R7] for two magnetic dipoles separated with a center-to-center distance of *R* as shown in Fig 2(b) of the Letter. We assume that the magnetostrictive layer is a random polycrystal so that we can neglect magnetocrystalline anisotropy.

The in-plane shape anisotropy produces the potential barrier between the two stable magnetization orientations along the easy axis (or y-axis). A compressive stress lowers this barrier by introducing stress anisotropy. We apply just the critical amount of stress necessary to "flatten" the shape anisotropy barrier, without inverting it to create a "well" instead of a "barrier". The potential profile however is not symmetric, but slightly tilted in one direction because of the dipole coupling term. This tilt causes the system to switch as it relaxes to the ground state. *The LLG equation is used to confirm the time to switch (e.g. for a 0.02 eV tilt we need ~100 ns to switch) and verify that the energy dissipation in the process is indeed equal to the energy of the "tilt" as shown in Fig S1. This is the case as loss in potential energy due to the "tilt" must be dissipated through the Gilbert damping before the magnetization settles to the ground state. Had this energy NOT been completely dissipated through damping, the kinetic energy of the system would build to a point where the system has no fault tolerance.*

The internal energy dissipated in the magnet during the magnetization rotation can be calculated as [R1] in the absence of thermal noise.

$$E_d(\tau) = \int_0^\tau -\left(\frac{dE_d}{dt}\right) dt = -\mu_0 \int_\Omega \vec{H}_{eff} \cdot \frac{d\vec{M}}{dt} d\Omega = \int_0^\tau \frac{\alpha \mu_0 \nu \Omega}{(1+\alpha^2)M_s} |\vec{H}_{eff}(t) \times \vec{M}(t)|^2 \, dt \quad \text{(S5)}$$

.

where $\tau$ is the time taken to switch.

While Equations (S1-S4) are adequate to describe the magnetization dynamics at 0 K, at room temperature (300 K), there is a random thermal torque that acts on the magnetization vector. This torque is $\vec{M}_i(t) \times \vec{h}(t)$ where $\vec{h}(t) = h_x(t)\hat{e}_x + h_y(t)\hat{e}_y + h_z(t)\hat{e}_z$ and each component $h_x(t)$, $h_y(t)$ and $h_z(t)$ is assumed to be isotropic, uncorrelated with other components, and has a Gaussian distribution given by:

$$h_i(t) = \sqrt{2\alpha kT / \gamma \mu_0 M_s \Omega(\Delta t)} * G(0,1) \quad \text{(S6)}.$$

where $T$ = temperature, $k$ = Boltzmann constant and $\Delta t$ = is related to the inverse of the attempt frequency for thermal torque to rotate the magnetization time and $G(0,1)$ is a Gaussian distribution with zero mean and unit standard deviation. The field $\vec{h}(t)$ is added vectorially to $\vec{H}_{eff}(t)$ in Equation (S1) to model the effect of thermal noise. We performed stochastic LLG simulations mentioned above in the manner of [R8] as we are interested in comparing a model nanomagnetic system (where each nanomagnet is treated as a macro-spin) with the analytical limit for the minimum energy dissipation for a given error probability.

To further illustrate how the out-of-plane distribution of magnetization hurts the $p_{error}$ in the model system and causes significant deviation from our analytical results under certain conditions, we plot magnetization distribution with respect to $\Phi$ (in-plane) and $\theta$ (out-of-plane) just prior to restoration of the barrier (withdrawal of stress). This is plotted at three temperatures: T=4.2 K (Fig S2), T=77K (Fig S3), T=300K (Fig S4). We also plot magnetization trajectories (Fig S5 a, b) that start close to $\Phi=0°$ (hard-axis) but can end up in either the $\Phi=90°$ (up) or $\Phi=-90°$ (down) position depending on the $\theta$ (out-of-plane orientation).

Now consider, the $\theta$ and $\phi$ distribution in Fig S3 (simulation at 77K). While the dipole is strong enough to ensure that the magnetization has switched close to the $\Phi=90°$ orientation, there is a small (but not negligible) probability that the magnetization is oriented, for example at $0<\Phi<\sim50°$. Now, half of these magnetization orientations would be oriented above the plane (torque due to demagnetizing field helps switching) and the other half oriented below the plane (torque due to demagnetizing field hurts switching). So, a small but significant fraction of the magnetizations will be oriented such that $\Phi>0$ and we would typically assume those will switch to the correct (up or $\Phi=90°$ state) but may actually switch back to the wrong orientation (up $\Phi=-90°$ state). This causes a decrease in the rate at which error probability decreases with increasing dipole strength particularly when we reach error limits ~1%. Since, this spread in magnetization is lower at 4.2K, lower $P_{error}$ is possible at lower temperatures. The spread is of course higher at 300 K leading to higher $p_{error}$ as shown in Fig S2.and Fig S4 respectively.

**FIGURE CAPTIONS**

S1. Comparison of dipole energy ($E_{tilt}$) and actual energy dissipated as calculated using LLG at 0K as a function of the distance separating the centers of the two nanomagnets.

S2. Magnetization distribution at 4.2 K in the (a) out-of-plane ($\theta$) (b) in-plane ($\Phi$) orientations.

S3. Magnetization distribution at 77 K in the (a) out-of-plane ($\theta$) (b) in-plane ($\Phi$) orientations.

S4. Magnetization distribution at 300 K in the (a) out-of-plane ($\theta$) (b) in-plane ($\Phi$) orientations.

S5. Simulation of phi (in-plane) magnetization dynamics (in the absence of thermal noise) for a magnetization that is initially oriented at $\Phi = 0$ (on the hard axis) but with:

(a) $\theta = 80°$ ( or 10° below the plane) and therefore switches to the wrong ($\Phi = -90°$) state.

(b) $\theta = 100°$ (or 10° above the plane) and therefore switches to the corrected ($\Phi = 90°$) state.

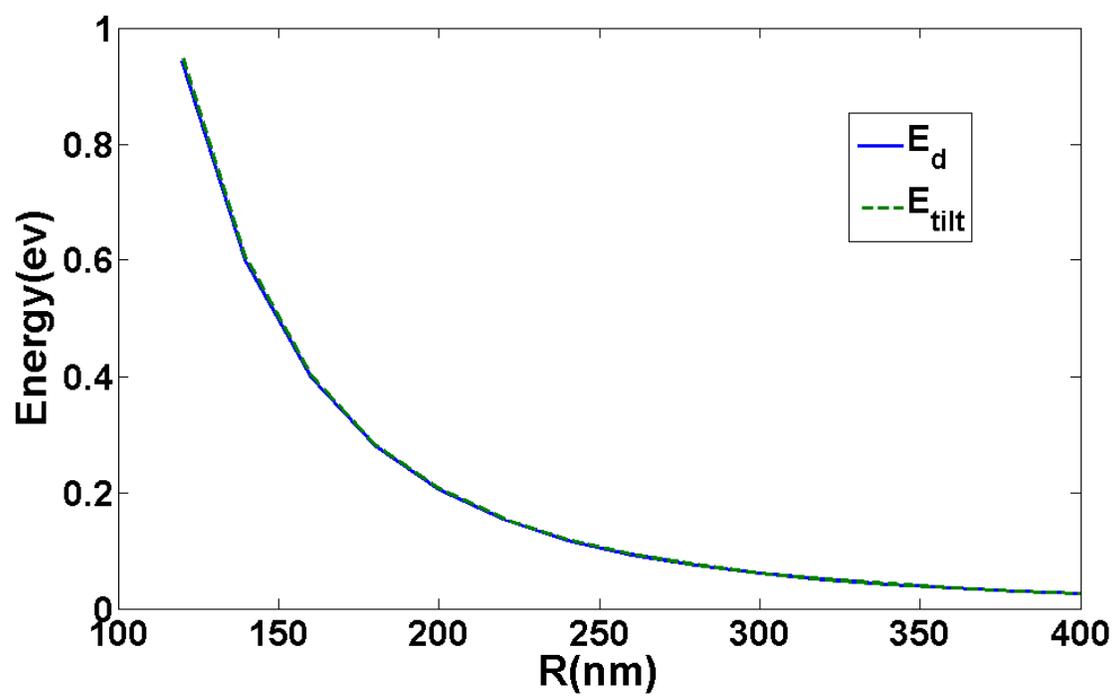

**Fig. S1**

(a)

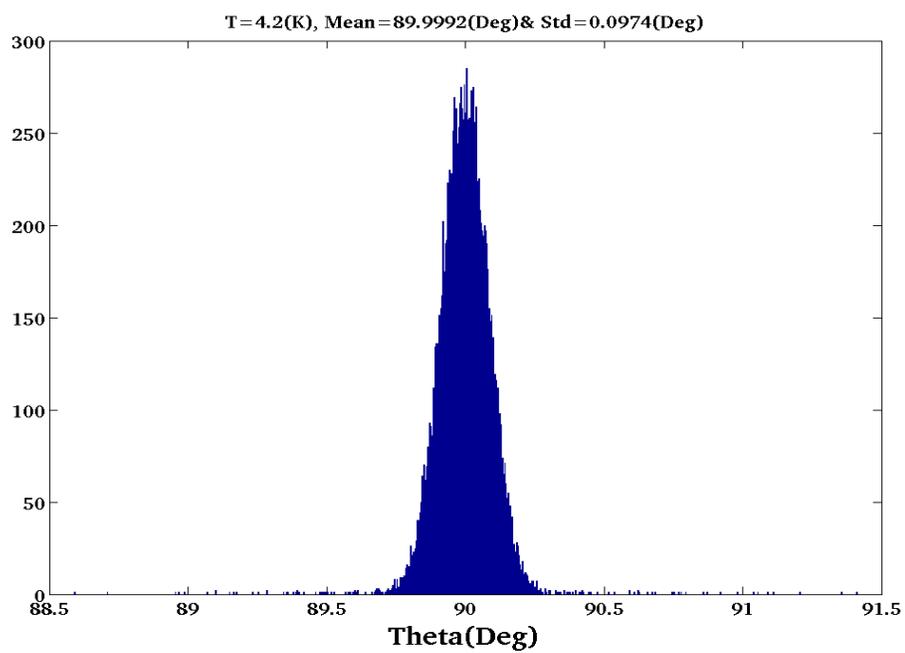

(b)

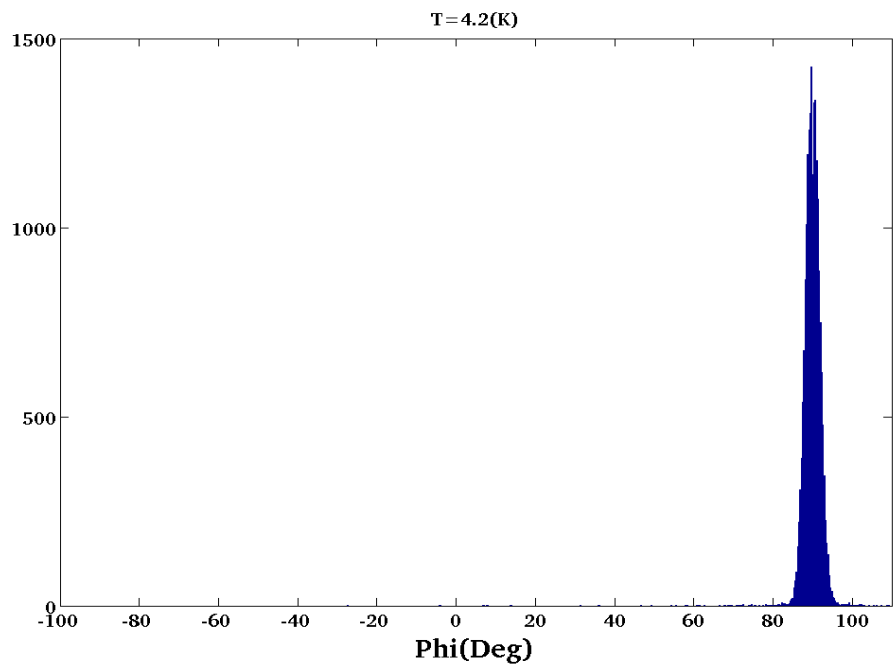

**Fig. S2**

(a)

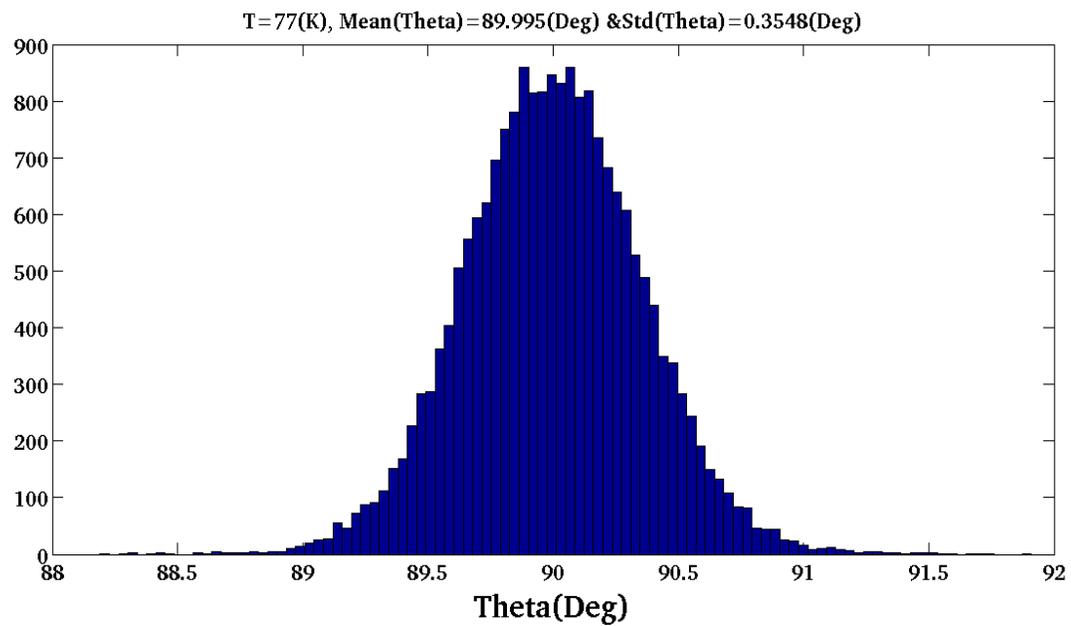

(b)

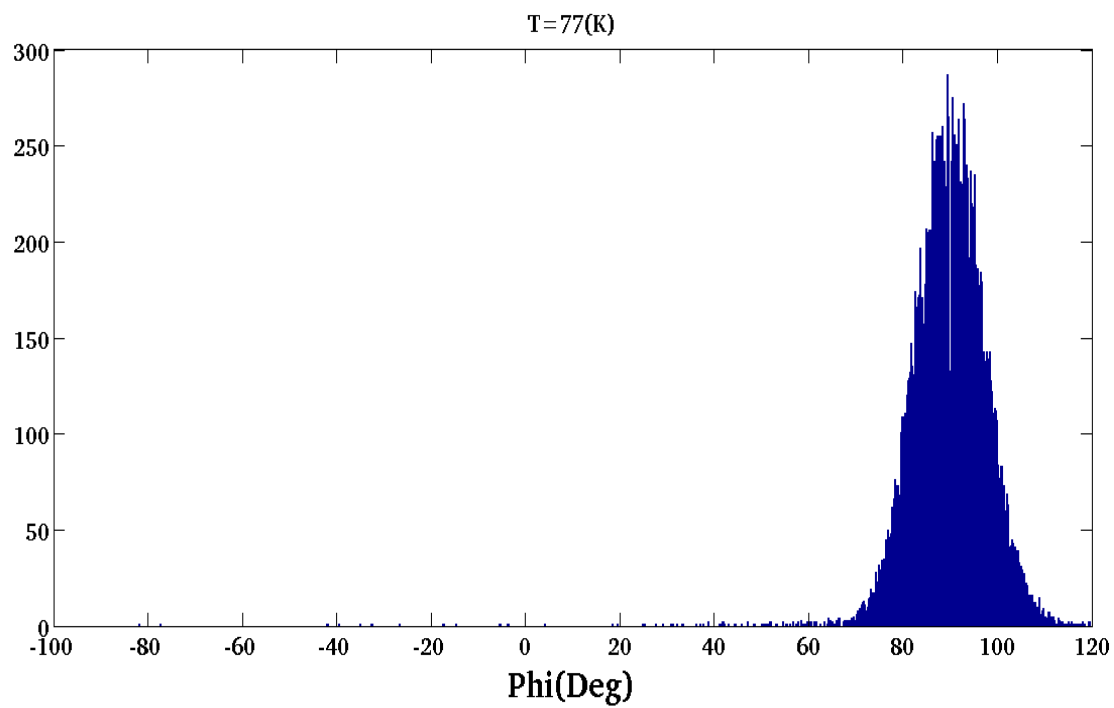

**Fig. S3**

**(a)**

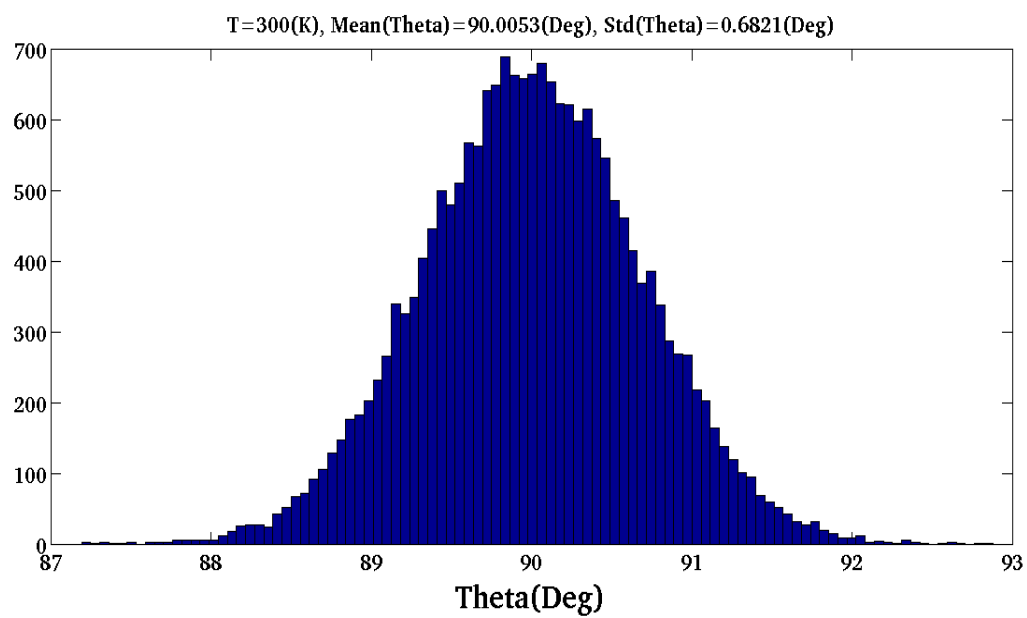

**(b)**

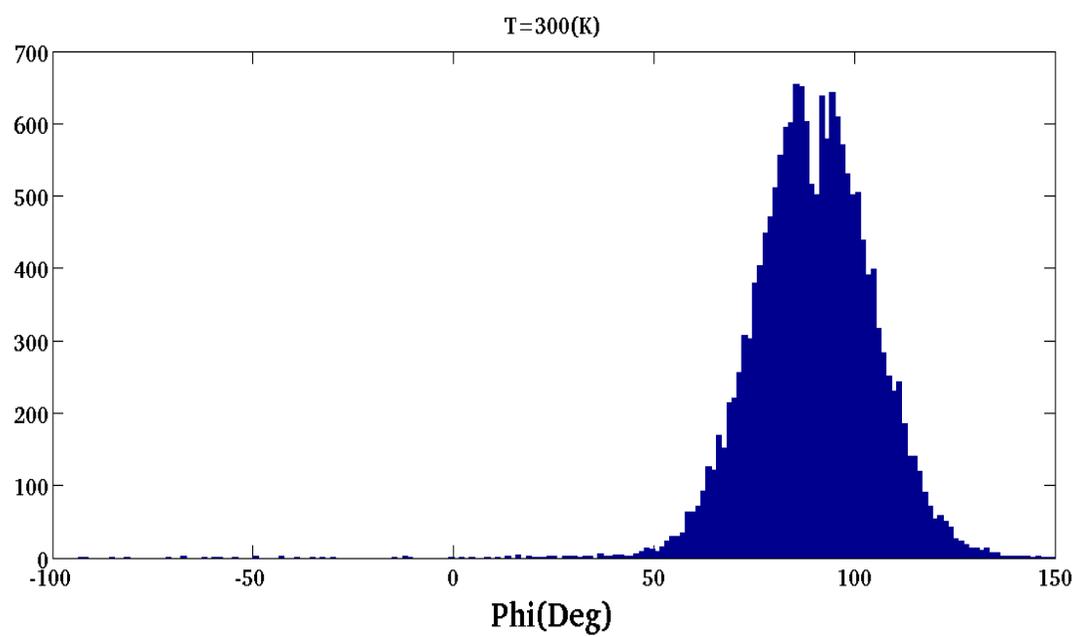

**Fig. S4**

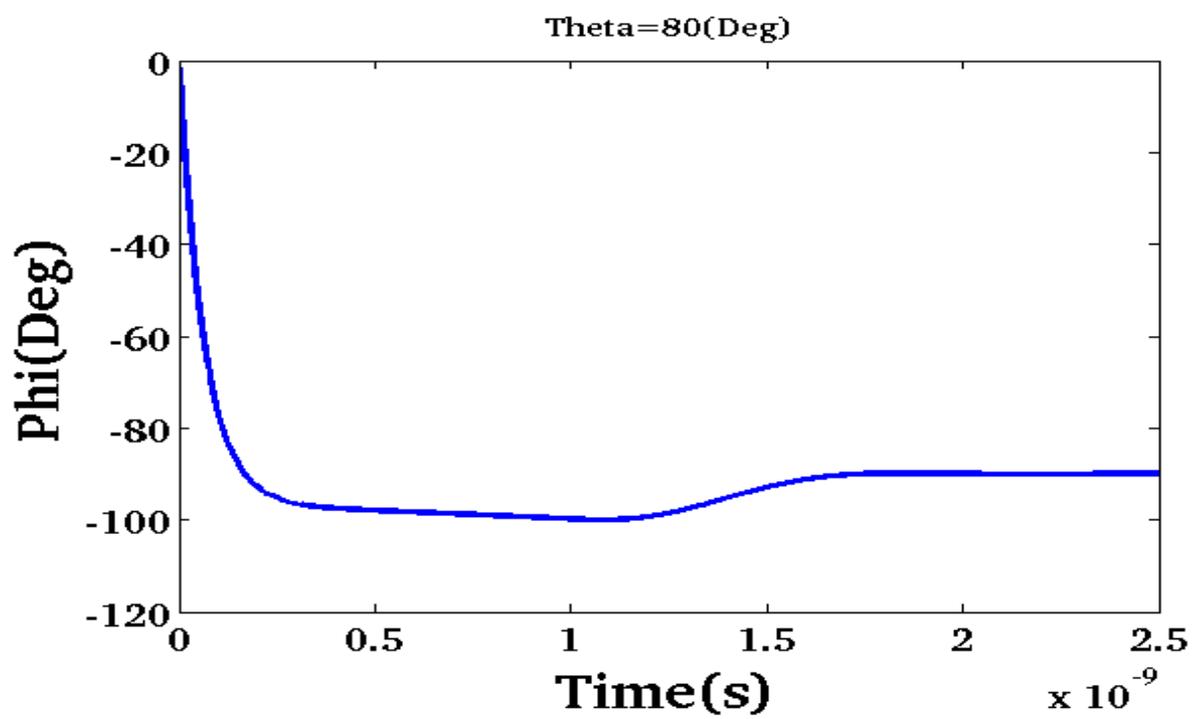
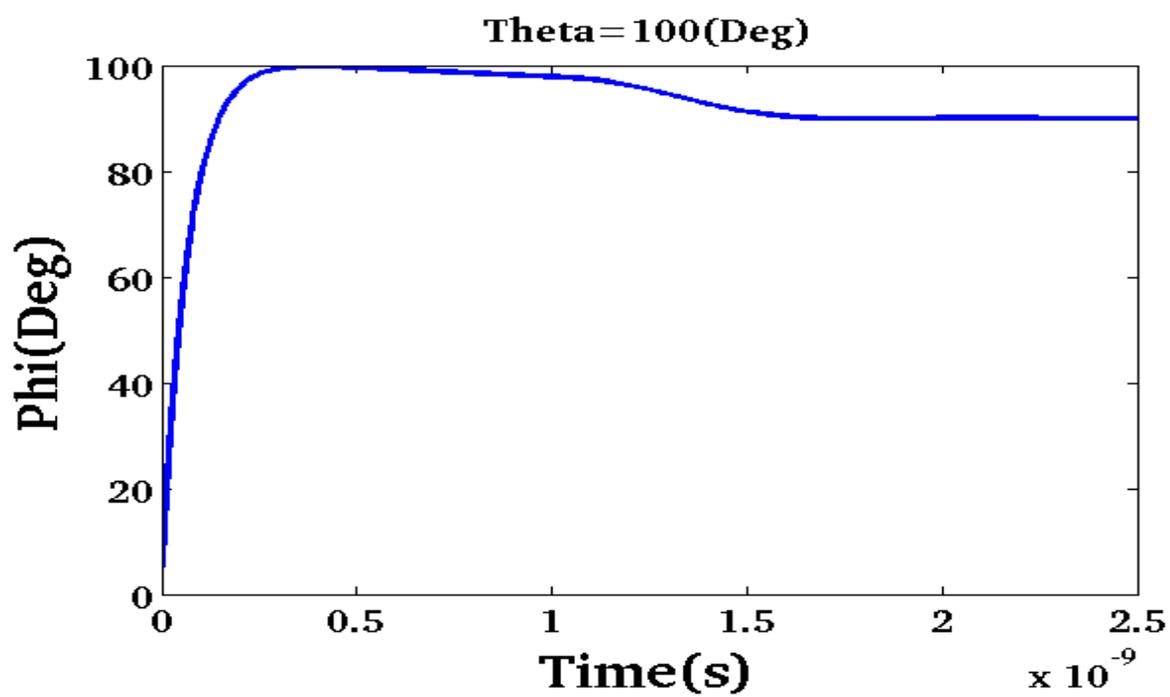

**Fig. S5**